\documentclass[sigconf]{acmart}

\acmConference[TEI '26]{The 20th International Conference on Tangible, Embedded, and Embodied Interaction}{March 8--11, 2026}{Chicago, USA}
\acmDOI{10.1145/3731459.3779050}

\usepackage{booktabs}
\usepackage{array}
\usepackage{graphicx}

\begin{document}

\title[Tangible Intangibles]{Tangible Intangibles: Exploring Embodied Emotion in Mixed Reality for Art Therapy}

\author{Mahsa Nasri}
\email{nasri.m@northeastern.edu}
\affiliation{%
  \institution{Northeastern University}
  \streetaddress{}
  \city{Boston}
  \state{MA}
  \country{USA}
  \postcode{}
}

\author{Mahnoosh Jahanian}
\email{jahanianmahnoosh01@gmail.com}
\affiliation{%
  \institution{Creative Technologist}
  \streetaddress{}
  \city{Dubai}
  \state{}
  \country{UAE}
  \postcode{}
}

\author{Wei Wu}
\email{wu.w4@northeastern.edu}
\affiliation{%
  \institution{Northeastern University}
  \streetaddress{}
  \city{Boston}
  \state{MA}
  \country{USA}
  \postcode{}
}

\author{Binyan Xu}
\email{xu.biny@northeastern.edu}
\affiliation{%
  \institution{Northeastern University}
  \streetaddress{}
  \city{Boston}
  \state{MA}
  \country{USA}
  \postcode{}
}
\author{Casper Harteveld}
\email{c.harteveld@northeastern.edu}
\affiliation{%
  \institution{Northeastern University}
  \streetaddress{}
  \city{Boston}
  \state{MA}
  \country{USA}
  \postcode{}
}

\begin{abstract}
This in-person studio explores how mixed reality (MR) and biometrics can make intangible emotional states tangible through embodied art practices. We begin with two well-established modalities, clay sculpting and free-form 2D drawing, to ground participants in somatic awareness and manual, reflective expression. Building on this baseline, we introduce an MR prototype that maps physiological signals (e.g., breath, heart rate variability, eye movement dynamics) to visual and spatial parameters (color saturation, pulsing, motion qualities), generating ''3D emotional artifacts.'' The full-day program balances theory (somatic psychology, embodied cognition, expressive biosignals), hands-on making, and comparative reflection to interrogate what analog and digital modalities respectively afford for awareness, expression, and meaning-making. Participants will (1) experience and compare analog and MR-based journaling of emotion; (2) prototype and critique mappings from biosignals to visual/spatial feedback; and (3) articulate design principles for trauma-informed, hybrid workflows that amplify interoceptive literacy without overwhelming the user. The expected contributions include a shared design vocabulary for biometric expressivity, a set of generative constraints for future TEI work on emotional archiving, and actionable insights into when automated translation supports or hinders embodied connection. 
\end{abstract}

\begin{teaserfigure}
  \includegraphics[width=\textwidth]{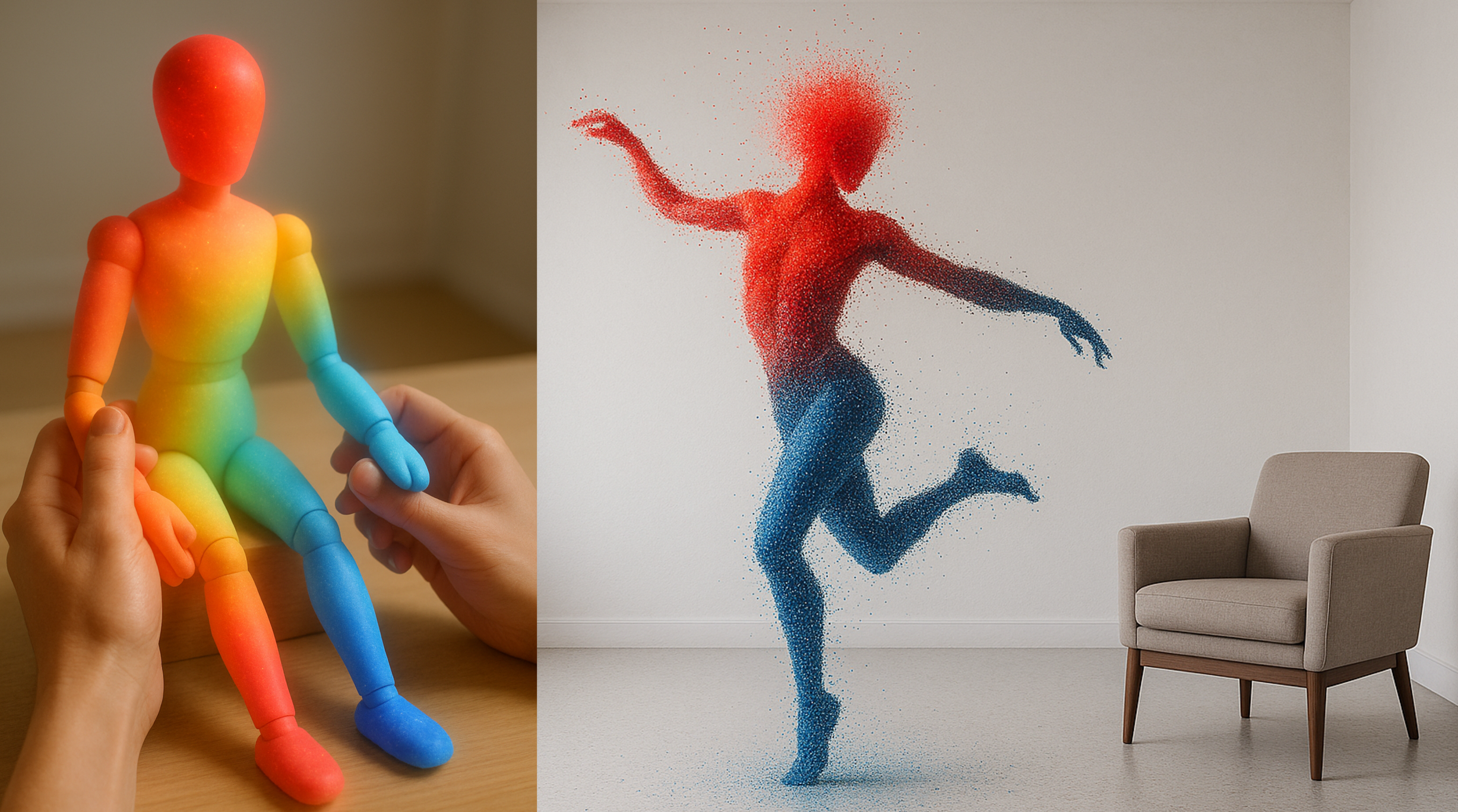}
  \caption{The right image: An MR environment where a particle figure dances in a white room, visualizing emotion through \textcolor{red}{red} and \textcolor{blue}{blue} energy fields. The left image: A mixed reality scene blending hands-on sculpting with glowing layers, mapping emotion onto a mannequin in real-time. Images generated by ChatGPT (DALL·E), by OpenAI.}
  \Description{An MR environment where a particle figure dances in a white room, visualizing emotion through red and blue energy fields. A mixed reality scene blending hands-on sculpting with glowing layers, mapping emotion onto a wooden mannequin in real-time. }
  \label{fig:teaser}
\end{teaserfigure}

\maketitle

\section{Motivation}
What is tangible?
Initial thoughts often center on objects that can be touched, held, or physically manipulated. But what about the tightness in your chest during anxiety? The weight of grief on your shoulders? The racing heart of excitement? These sensations are intensely felt; they have texture, pressure, temperature, rhythm, yet they are often considered intangible. We cannot pick them up, examine them from all sides, or place them on a shelf to revisit later. This studio proposal explores how to make the intangible tangible. This question sits at the heart of art therapy, somatic psychology, and embodied cognition.  

When a therapist asks you to ''sculpt your depression in clay,'' they are attempting to externalize what lives only inside. When a dancer moves through grief, they are giving form to formlessness. When a patient tracks their panic attacks in a journal, they are trying to see what usually remains invisible. Tangible Intangibles explores this ancient practice through contemporary biometric interfaces: What happens when your breath rhythm becomes color saturation? When your eye movements choreograph an avatar? When does your heartbeat determine the brightness of a joint?

We propose that emotions, though intangible, leave tangible traces in the body—respiratory patterns, cardiac rhythms, muscular tensions, and gaze behaviors. Mixed reality, paired with biometric sensors, can intercept these traces and transform them into material artifacts: 3D sculptures that can be rotated, revisited, and arranged into galleries of emotional memory.

This workshop begins with art therapy practice, then explores how emerging technologies might expand its possibilities. Participants will experience two pathways for expressing emotion through form:

\begin{itemize}
    \item \textbf{Traditional Art Therapy (Clay + Manual Color):} Participants engage with established somatic art therapy techniques, guided body scans, followed by sculpting and expressing emotional states. Through tactile manipulation and conscious color selection, they externalize felt sensations into material form. This grounds us in the embodied, reflective process that has served therapeutic practice for decades.
    
    \item \textbf{Mixed Reality-Enhanced Exploration (Movement + Biosignals):} Building on the body awareness developed through traditional practice, participants explore what Mixed Reality's (MR) unique affordances might add: \textit{What if your breath could directly modulate color without conscious choice? What if eye movements could reveal patterns of activation you don't consciously notice? What if movement over time could be captured as spatial traces?} Here, biosensors intercept the body's involuntary signals, translating them into visual feedback and persistent 3D artifacts.
\end{itemize}

Rather than comparing which approach is more "optimal," the studio asks: \textit{What can MR do that clay cannot? What does clay offer that MR loses? How might we design hybrid systems that combine the best of both?} Participants experience traditional art therapy first to establish somatic literacy, then experiment with how biometric interfaces might expand, enhance, or transform these practices—revealing emotional patterns that remain invisible in traditional media, enabling new forms of emotional journaling over time, or creating accessible alternatives for those intimidated by physical art-making.

The goal is generative: to prototype new possibilities for art therapy in the digital age while remaining grounded in the body's wisdom.


By experiencing both, participants investigate when automated biometric translation is valuable. When does manual interpretation create deeper connection? What hybrid approaches might combine real-time biofeedback with reflective material choice?

By experiencing both modalities, participants will investigate fundamental questions:

\begin{itemize}
    \item What can MR do that traditional art therapy cannot? What does clay offer that digital media loses?
    \item How might biosensors enhance emotional awareness? When might they distance us from embodied experience?
    \item What makes an emotional artifact meaningful—the process of making it, the ability to touch it, or the capacity to revisit it over time?
    \item How might we design hybrid systems that combine real-time biofeedback with reflective material choice?
\end{itemize}

Rather than optimizing emotional regulation, this workshop treats technology as a creative medium to explore intangible experience. Through guided meditation, free-form movement, sculptural, and collaborative prototyping, participants will imagine new possibilities for art therapy. We are exploring systems that honor the complexity of emotion, that amplify somatic awareness without overwhelming it, that create records of emotional experience without reducing feelings to data points.

The goal is not to solve the problem of intangibility, but to explore its generative potential: the space between felt experience and visual form, between the body that knows and the artifact that remembers, between what moves through us and what we can hold in our hands or revisit on our phones.

\section{Grounding in Theory}
The workshop draws from three intersecting theoretical grounds.
\subsection{Somatic Psychology and Trauma-Informed Design}
Following Peter Levine's \cite{levine2010} Somatic Experiencing and van der Kolk's \cite{vanderkolk2014} The Body Keeps the Score, emotions are not merely cognitive events but embodied states held in posture, breath, and movement. Making these states visible through biometric visualization enables what Levine calls ''pendulatio'' \cite{levine2010}, the oscillation between activation and regulation that characterizes emotional processing.

\subsection{Embodied Cognition and Interoceptive Awareness}
Based on Varela's \cite{varela1991} Embodied Mind and Craig's \cite{craig2009} interoceptive neuroscience, self-awareness emerges from the body's ability to sense its own internal states. Biometric interfaces that reflect back to users patterns of breathing, heart rate, and eye movements create an external mirror for internal proprioception, strengthening the feedback loop between body and conscious awareness.

\subsection{Digital Phenotyping and Expressive Biosignals}
Drawing on affective computing \cite{picard1997}, eye-tracking research \cite{lim2020emotion, nasri2024exploring}, and the quantified self movement, the workshop treats biosignals not as objective measurements but as expressive material—raw data that becomes meaningful through personal interpretation and aesthetic transformation. Like Vallgårda's \cite{vallgarda2014} material speculation, we ask: what does it mean to sculpt with your breath, paint with your heartbeat, choreograph with your gaze?

\section{The Role of XR}
In Tangible Intangibles, mixed reality functions not as entertainment or training but as a biometric mirror, a space where invisible physiological states become visible, manipulable, and memorable. Traditional emotional tracking apps reduce feelings to buttons and scales; they ask users to translate embodied experience into discrete categories. Our approach inverts this: the body speaks first, through breath rhythm, pupil dilation, and movement quality. MR becomes the translator, rendering these autonomic signals as color, form, and spatial trace. The system creates productive friction between voluntary action (choosing colors, moving limbs) and involuntary response (heart rate variability, saccadic movements). This dialogue mirrors the relationship between conscious and unconscious emotional processing.

\subsection{Experience Design}
Participants begin with a guided body scan meditation in MR, establishing interoceptive awareness, noticing where emotions are experienced in the body as sensation, tension, or energy. Transitioning to mixed reality, they encounter a simple articulated avatar representing their body.  They first make conscious choices: selecting colors for specific joints based on what they noticed during the body scan, perhaps blue for calm shoulders, \textcolor{red}{red} for an anxious chest, yellow for energized hands. Then, they begin to move freely. As they move, the avatar mirrors their gestures. Still, biosensors add a layer of involuntary expression: breathing depth modulates color intensity, heart rate variability affects pulsing patterns, and eye movement speed (saccades) influences the quality of motion—smooth and flowing when regulated, fragmented and staccato when processing activation.

When rapid, fragmented eye movements accelerate the avatar's motion—reflecting cognitive processing of past trauma or future anxiety—participants observe their own nervous system activation in real-time. When slow, deep breathing softens colors and smooths movement, they experience biofeedback of self-regulation.

The experience concludes with the system capturing the movement session as a 3D sculpture, a frozen moment preserving the pose, color palette, and movement traces. This artifact is stored in the cloud and can be viewed later on participants' phones, serving as a timestamp: "This is what my emotion looked like in my body on this day."

\textbf{Co-Design Process:} While we provide a working prototype to ground the exploration, the studio explicitly invites participants to reimagine, critique, and expand this design. Through hands-on experimentation and group reflection, participants will collaboratively generate alternative mappings (e.g., what if temperature modulated color instead of breath? What if the avatar could fracture or merge?) and propose hybrid analog-digital workflows, and sketch future possibilities for biometric art therapy systems. The prototype serves as a provocation, not a prescription, a starting point for collective inquiry into what emotional interfaces could become.

By capturing these moments as persistent 3D artifacts, MR transforms ephemeral emotional states into an embodied archive. Over time, the collection reveals patterns—recurring postures, evolving color palettes, shifting movement qualities—that would otherwise remain invisible. A person might notice: "Every time I process work stress, I raise my right shoulder and turn \textcolor{red}{red}. When I'm grieving, I curl inward and fade to \textcolor{blue}{blue}."

The virtual and the tangible do not oppose one another but form a biometric feedback loop: body → sensors → visualization → awareness → body. MR amplifies what the body already knows but cannot easily see, creating new possibilities for emotional self-knowledge and therapeutic practice.

\section{Learning Goals}

\begin{enumerate}
   \item Learn to express themselves through traditional art therapy and mindfulness. 
    \item Develop skills to prototype systems that translate physiological signals into meaningful visual/spatial feedback for self-awareness and emotional processing.

    \item Through experiencing both digital (MR + biometrics) and analog (clay + manual color) approaches, participants develop critical insight into what each modality affords for emotional expression.
    \item Experiential understanding of how emotions manifest as physiological patterns (breath, heart rate, eye movement) and can be witnessed through biometric visualization.
    \item Understanding how persistent digital artifacts (3D emotional sculptures) can function as reflective tools for tracking emotional patterns over time.

\end{enumerate}

Participants will cultivate biometric empathy, a sensitivity to how invisible physiological states can be made tangible, witnessed, and archived.

\section{Studio Schedule}
The studio is an \textit{in-person} event that begins with two well-established art therapy practices, clay sculpting and free-form 2D drawing. It then proceeds to testing mixed reality prototypes that reimagine how these techniques might appear and feel in new media. The detailed schedule is provided in Table~\ref{tab:schedule}.
The schedule balances hands-on traditional practice, conceptual framing, and speculative prototyping to reimagine art therapy methods that can \textit{Tangible Intangible}.

\begin{table*}[ht]
\centering
\caption{Full-day studio structure and activities overview.}
\label{tab:schedule}
\small
\begin{tabular}{p{2.2cm} p{3.2cm} p{7.8cm}}
\toprule
\textbf{Time} & \textbf{Activity} & \textbf{Description} \\
\midrule
09:00–09:30 & Arrival \& Somatic Grounding & Guided body scan meditation introducing interoceptive awareness. Participants notice breath, heart, tension, and emotion in the body, establishing baseline somatic attention. \\
\addlinespace
09:30–10:00 & Theoretical Context & Mini-presentations on somatic psychology, biometric interfaces as emotional mirrors, and digital phenotyping. Brief demo of the MR biometric journaling system. \\
\addlinespace
10:00– 11:00 &  Experience Session: Clay \& Manual Color &Participants sculpt with physical clay on mannequin forms, selecting and applying colors guided by body-scan awareness. They also complete free-form 2D art therapy with paper and pencil. A photographic record is captured for comparison.  \\
\addlinespace
11:00–11:15 & Break \& Informal Reflection & Coffee/tea break with discussion prompt: "What surprised you about seeing your emotion embodied?" \\
\addlinespace
11:30–12:30 & Experience Session: MR Biometric Journaling & Participants (in pairs, rotating) experience the full system: body scan meditation, an MR avatar with joint color selection. They express free movement while biometrics modulate color/speed into 3D artifact generation. Non-participating members observe and take notes. 
\\
\addlinespace
12:30–13:30 & Comparative Reflection \& Prototyping & Small groups discuss: What did digital vs. analog modalities afford? Where did you feel more connected to emotion? Sketch ideas for hybrid or alternative biometric emotion journaling systems. \\
\addlinespace
13:30–14:30 & Lunch Break & Informal Discussions with participants. \\
14:30–16:00 & Gallery Walk \& Closing Circle & Display both MR artifacts (via tablets/phones showing the models) and clay sculptures. Participants share one insight. Co-create a visual "emotion mapping" poster synthesizing the day's discoveries. \\

\bottomrule
\end{tabular}
\end{table*}

\section{Expected Contributions}
This workshop contributes to TEI's discourse by expanding biometric interfaces beyond performance optimization to explore biosignals as material for self-knowledge and emotional expression. By running parallel analog clay and digital MR emotion journaling sessions, we introduce comparative embodiment research that generates lived-experience data on what each modality affords for somatic awareness. The 3D artifact archive system prototypes emotion journaling as a design space for temporal emotional tracking over extended periods. We center trauma-informed design in MR by grounding the experience in body-scan meditation and incorporating EMDR principles, demonstrating how immersive technology can support emotional safety rather than overwhelm users. Participants leave with a design vocabulary—including biometric color mapping, movement quality modulation, and embodied archiving that applies to mental health, therapeutic MR, wearables, and beyond.

\begin{acks}
We thank Ghost Lab at Northeastern University for equipment and support.
\end{acks}

\bibliographystyle{ACM-Reference-Format}
\bibliography{references}

\end{document}